\documentclass[twocolumn,prd,aps,superscriptaddress,preprintnumbers,tightenlines,showpacs,nofootinbib,eqsecnum,amsfonts,amsmath]{revtex4-1}
\usepackage{subfigure}
\usepackage[dvipsnames, usenames]{xcolor}
\definecolor{linkcolor}{rgb}{0.0,0.3,0.5}
\usepackage[unicode, colorlinks=true, linkcolor=linkcolor, citecolor=linkcolor, filecolor=linkcolor,urlcolor=linkcolor, pdfusetitle]{hyperref}
\usepackage[all]{hypcap}
\usepackage[T1]{fontenc}
\usepackage[utf8]{inputenc}
\usepackage{orcidlink}
\usepackage{overpic,makecell,multirow} 
\usepackage{xtab,afterpage,longtable}
\usepackage[normalem]{ulem}
\usepackage{placeins}
\usepackage{physics}

\newcommand{\bpop}{\textsc{B-Pop}~}

\begin{document}

\title{Comparing astrophysical models to gravitational-wave data in the observable space}

\author{Alexandre Toubiana\,\orcidlink{0000-0002-2685-1538}}
\email{alexandre.toubiana@unimib.it}
\affiliation{Dipartimento di Fisica ``G. Occhialini'', Università di Milano-Bicocca, Milano, Italy}
\affiliation{INFN, Sezione di Milano-Bicocca, Piazza della Scienza 3, 20126 Milano, Italy}

\author{Davide Gerosa\,\orcidlink{0000-0002-0933-3579}}
\affiliation{Dipartimento di Fisica ``G. Occhialini'', Università di Milano-Bicocca, Milano, Italy}
\affiliation{INFN, Sezione di Milano-Bicocca, Piazza della Scienza 3, 20126 Milano, Italy}

\author{Matthew Mould\,\orcidlink{0000-0001-5460-2910}}
\affiliation{Nottingham Centre of Gravity \& School of Mathematical Sciences, University of Nottingham,\\University Park, Nottingham, NG7 2RD, United Kingdom}
\affiliation{LIGO Laboratory, Massachusetts Institute of Technology, Cambridge, MA 02139, USA}
\affiliation{Kavli Institute for Astrophysics and Space Research, Massachusetts Institute of Technology, Cambridge, MA 02139, USA}

\author{Stefano Rinaldi\,\orcidlink{0000-0001-5799-4155}}
\affiliation{Institut für Theoretische Astrophysik, Zentrum für Astronomie, Universität Heidelberg, Albert-Ueberle-Str. 2, 69120 Heidelberg, Germany}

\author{\\ Manuel Arca Sedda\, \orcidlink{0000-0002-3987-0519}}
\affiliation{Gran Sasso Science Institute (GSSI), L'Aquila (Italy),
            Viale Francesco Crispi 7}
\affiliation{INFN, Laboratori Nazionali del Gran Sasso, 67100 Assergi, Italy}

\author{Tristan Bruel\,\orcidlink{0000-0002-1789-7876}}
\affiliation{Dipartimento di Fisica ``G. Occhialini'', Università di Milano-Bicocca, Milano, Italy}
\affiliation{INFN, Sezione di Milano-Bicocca, Piazza della Scienza 3, 20126 Milano, Italy}

\author{Riccardo Buscicchio\, \orcidlink{0000-0002-7387-6754}}
\affiliation{Dipartimento di Fisica ``G. Occhialini'', Università di Milano-Bicocca, Milano, Italy}
\affiliation{INFN, Sezione di Milano-Bicocca, Piazza della Scienza 3, 20126 Milano, Italy}
\affiliation{Institute for Gravitational Wave Astronomy \& School of Physics and Astronomy, University of Birmingham, Birmingham, B15 2TT, UK}

\author{Jonathan Gair\,\orcidlink{0000-0002-1671-3668}}
\affiliation{Max Planck Institute for Gravitational Physics (Albert Einstein Institute), Am M\"{u}hlenberg 1, 14476 Potsdam, Germany}

\author{\\Lavinia Paiella\, \orcidlink{0009-0001-7605-991X}}
\affiliation{Gran Sasso Science Institute (GSSI), L'Aquila (Italy),
            Viale Francesco Crispi 7}
\affiliation{INFN, Laboratori Nazionali del Gran Sasso, 67100 Assergi, Italy}

\author{Filippo Santoliquido\, \orcidlink{0000-0003-3752-1400}}
\affiliation{Gran Sasso Science Institute (GSSI), L'Aquila (Italy),
            Viale Francesco Crispi 7}
\affiliation{INFN, Laboratori Nazionali del Gran Sasso, 67100 Assergi, Italy}

\author{Rodrigo Tenorio\,\orcidlink{0000-0002-3582-2587}}
\affiliation{Dipartimento di Fisica ``G. Occhialini'', Università di Milano-Bicocca, Milano, Italy}
\affiliation{INFN, Sezione di Milano-Bicocca, Piazza della Scienza 3, 20126 Milano, Italy}

\author{Cristiano Ugolini\, \orcidlink{0009-0005-9890-4722} }
\affiliation{Gran Sasso Science Institute (GSSI), L'Aquila (Italy),
            Viale Francesco Crispi 7}
\affiliation{INFN, Laboratori Nazionali del Gran Sasso, 67100 Assergi, Italy}

\begin{abstract}
    Comparing population-synthesis models to the results of hierarchical Bayesian inference in gravitational-wave astronomy requires a careful understanding of the domain of validity of the models fitted to data. This comparison is usually done using the inferred astrophysical distribution: from the data that were collected, one deconvolves selection effects to reconstruct the generating population distribution. In this paper, we demonstrate the benefits of instead comparing observable populations directly. In this approach, the domain of validity of the models is trivially respected, such that only the relevant parameter space regions as predicted by the astrophysical models of interest contribute to the comparison. With this in mind, it can be useful to fit the observed population directly, rather than effectively deconvolving the selection effects only to fold them back in when reconstructing the observable population. We clarify that unbiased inference of the observable compact-binary population is indeed possible. Crucially, this approach still requires incorporating selection effects, but in a manner that differs from the standard implementation. We apply our observable-space reconstruction to LIGO-Virgo-KAGRA data from their third observing run and illustrate its potential by comparing the results to the predictions of a fiducial population-synthesis model.
\end{abstract}

\maketitle

\section{Introduction}
The growing number of gravitational-wave (GW) detections by LIGO, Virgo, and KAGRA (LVK)~\cite{LIGOScientific:2014pky, VIRGO:2014yos, KAGRA:2020tym} now enables detailed studies of the source population. The third Gravitational-Wave Transient Catalog (GWTC-3)~\cite{KAGRA:2021vkt} allowed us to infer the binary black hole (BBH) mass distribution, measure merger rates out to redshift $z \sim 1$,~and begin probing parameter correlations~\cite{Callister:2024cdx}, offering key insights into their formation~\cite{Mandel:2018hfr,Mapelli:2021taw}. Comparing population inferences with population-synthesis predictions is essential to identify formation channels and place stellar-mass black holes in their broader astrophysical context.

GW population inference is carried out using a hierarchical Bayesian framework~\cite{Mandel:2018mve,Vitale:2020aaz}. Given a model for the source population that depends on a set of parameters—commonly referred to as hyperparameters—this formalism enables their inference while properly accounting for both measurement uncertainties and selection~effects. 

Astrophysically informed population models directly constrain the physical processes driving binary formation (e.g., Refs.~\cite{Zevin:2020gbd,Wong:2020ise,Bouffanais:2020qds,Franciolini:2021tla,Mould:2022ccw,Cheng:2023ddt,Colloms:2025hib,Plunkett:2025mjr}). However, they require efficient interpolation techniques to evaluate the merger rate across source parameters and hyperparameters during hierarchical inference (but see Refs.~\cite{Leyde:2023iof,Jiang:2025jxt} for alternative simulation-based approaches). Moreover, such models allow for limited flexibility in representing the properties of the source population and make it harder for unpredicted features to emerge from observations.

For these reasons, parametric (e.g., \cite{Talbot:2018cva,Callister:2021fpo,Adamcewicz:2022hce,Biscoveanu:2022qac,Franciolini:2022iaa,Wang:2022gnx,Li:2023yyt,Pierra:2024fbl,Gennari:2025nho,Hussain:2024qzl}) and non-parametric (e.g., \cite{Rinaldi:2021bhm,Tiwari:2021yvr,Sadiq:2021fin,Edelman:2022ydv,Ruhe:2022ddi,Farah:2023vsc,Callister:2023tgi,Toubiana:2023egi, Heinzel:2024jlc, Tenorio:2025nyt}) approaches have been commonly adopted for modeling the astrophysical population. Parametric models offer simplicity but can extrapolate poorly beyond the region constrained by data, potentially leading to erroneous reconstruction of the population outside the informative portion of parameter space, which can in turn produce misleading conclusions when comparing to predictions from astrophysical simulations, as we illustrate in this paper. Non-parametric models are more flexible but often rely heavily on built-in priors where data are sparse, raising questions about inference in regions where detector sensitivity is limited.

One could instead compare GW population fits and astrophysical models directly in the observable space. To obtain the observable distribution, the current approach is to first infer the astrophysical distribution using the standard formalism~\cite{Mandel:2018mve,Vitale:2020aaz} and then apply selection effects to it, effectively deconvolving selection effects only to fold them back in. Can one instead compute the observable population directly? %This will still requires hierarchical modeling to deconvolve measurement uncertainties and, moreover, 
Reference~\cite{Essick:2023upv} showed that accounting for selection effects only after inference, which might appear to be the natural procedure, actually leads to biased results, but did not propose an approach for direct inference in observable space. 

In this work, we show that it is possible to perform unbiased inference on the observable population without reconstructing the astrophysical distribution first, provided that selection effects are consistently incorporated during inference. We apply this method to the events observed during the LVK’s third observing run (O3) and, using the population-synthesis model of Ref.~\cite{2023MNRAS.520.5259A} as an example, we demonstrate concretely how performing the comparison directly in observable space can serve as a valuable alternative approach for astrophysical interpretation.

\section{Formalism}\label{sec:formalism}

We first recall the standard formalism for inferring the astrophysical distribution, define the observable population based on it, and then show how the standard formalism can be modified to infer the observable population directly.

\subsection{From astrophysical to observable populations}

We denote by $\theta$ the set of parameters characterizing an astrophysical event and by $\frac{\dd{N_\mathrm{A}}}{\dd{\theta}}(\Lambda)$ the differential astrophysical number of events, which depends on a set of hyperparameters $\Lambda$ that we aim to infer. The astrophysical population prior is defined through $\frac{\dd{N_\mathrm{A}}}{\dd{\theta}}(\Lambda) = N_\mathrm{A} p_\mathrm{A}(\theta|\Lambda)$,
where $N_{\rm A}$ is the expected astrophysical number of events.
The most common formalism for population inference in GW astronomy is to constrain the hyperparameters $\Lambda$ using only detected events~\cite{Mandel:2018mve,Vitale:2020aaz} (but approaches using information from marginal triggers \cite{2019MNRAS.484.4008G,2020PhRvD.102l3022R} and the stochastic background produced by quiet BBHs~\cite{Callister:2020arv,Smith:2020lkj,Bers:2025tei,Cousins:2025bas,Ferraiuolo:2025evh,Giarda:2025ouf} have also been proposed).
The population likelihood for observing $N_\mathrm{E}$ events $\{d\}=\{d_1,...,d_{N_\mathrm{E}}\}$ is given by~\cite{Mandel:2018mve,Vitale:2020aaz} 
\begin{align}
p(\{d\}|\Lambda) \propto e^{-N_{\rm A}p({\rm det}|\Lambda)}\prod_{i=1}^{N_\mathrm{E}}\int {\rm d}\theta_i \  p(d_i|\theta_i)\frac{{\rm d}N_{\rm A}}{{\rm d}\theta_i}(\Lambda)\, .\label{eq:std_diff_rate}
\end{align}
The selection function is defined as 
\begin{align}
p({\rm det}|\Lambda) =
\int \dd{\theta} p({\rm det}|\theta) p_{\rm A}(\theta|\Lambda)
\, ,
\end{align} 
where 
\begin{align}
p({\rm det}|\theta)=\int_{d>{\rm threshold}} \dd{d}\, p(d|\theta)\label{eq:pdet}
\end{align}
is the single-parameter detection probability and $ p(d|\theta)$ is the usual GW likelihood. % of detectable data $d$. 
%In the current modus operandi of LVK,
Detection is determined by applying a threshold to a ranking statistic—such as the false-alarm rate—evaluated on noisy data $d$. This statistic is a deterministic function of a given signal-plus-noise realization.
As a result, the detectability of a hypothetical signal with parameters $\theta$ without considering a specific data realization is inherently a probabilistic quantity.

The observable population is defined as the astrophysical population conditioned on detection: 
\begin{align}
p_{\rm O}(\theta|\Lambda) := p_{\mathrm{A}}(\theta|\mathrm{det}, \Lambda) = \frac{ p({\rm det}|\theta) p_{\rm A}(\theta|\Lambda) }{ p({\rm det}|\Lambda) }\, \label{eq:def_pobs}.
\end{align}
Similarly, the observable differential number of events is given by 
\begin{align}
    \frac{{\rm d}N_{\rm O}}{{\rm d}\theta}(\Lambda)= p({\rm det}|\theta)\frac{{\rm d}N_{\rm A}}{{\rm d}\theta}(\Lambda)=N_{\rm O}p_{\rm O}(\theta|\Lambda), \label{eq:obs_rate}
\end{align}
and the expected number of observable events is $N_{\rm O}=p({\rm det}|\Lambda)N_{\rm A}$. 

In practice, $p({\rm det}|\theta)$ can be estimated via Monte Carlo integration of Eq.~\eqref{eq:pdet} by injecting signals with parameters $\theta$ into simulated (or real) detector noise and averaging the detection statistic over noise realizations. As shown in Refs.~\cite{Gerosa:2020pgy, Talbot:2020oeu, Callister:2024qyq}, emulators can accelerate this process by efficiently predicting $p({\rm det}|\theta)$ across parameter space. Due to the high dimensionality, analytically computing the observable population from the astrophysical one is impractical. Instead, we sample from the astrophysical distribution, evaluate $p({\rm det}|\theta)$ for each sample, and reweight accordingly—an approach we use to obtain the LVK and model predictions in observable space; see~Fig.~\ref{fig:plots}. We now discuss how the observable population can be directly inferred, without resorting to the two-step procedure in which selection effects are first deconvolved and then folded back in.

\subsection{Direct inference on the observable population}\label{sec:obs_pop}

Using the definition of the observable differential number of events from Eq.~\eqref{eq:obs_rate}, one can rewrite Eq.~\eqref{eq:std_diff_rate}  as
\begin{align}
p(\{d\}|\Lambda) &\propto e^{-N_{\rm O}} \prod_i^{N_\mathrm{E}}\int {\rm d}\theta_i \frac{p(d_i|\theta_i)}{p({\rm det}|\theta_i)}\frac{{\rm d}N_{\rm O}}{{\rm d}\theta_i}(\Lambda) .\label{eq:convert}
\end{align}
This expression allows inferring the observable population directly, without reconstructing the astrophysical one~first.

Our approach differs from the ``inferring the detected distribution'' formalism of Ref.~\cite{Essick:2023upv}, which discussed a flawed method that had been used within the GW community to infer the observable population in order to demonstrate that this method led to biased results. There are two key differences:  
\begin{enumerate}
    \item Our approach correctly treats detection as dependent directly only on observed data $d$ (regardless of whether that dependence is deterministic or not, though is taken here to be so, in line with the usual approach), not on source properties $\theta$ [see Eq.~\eqref{eq:pdet}].
    \item The single-event likelihood appears divided by the detection probability, the $p(d_i|\theta_i)/p({\rm det}|\theta_i)$ terms, which effectively acts as a renormalization of the single-event likelihood over the space of observable data.\footnote{Some previous works (e.g., Ref.~\cite{Rinaldi:2024fgq}) neglected the $p({\rm det}|\theta_i)$ term. This approximation holds only in the regime in which the likelihood $p(d_i|\theta_i)$ changes more rapidly than the selection function. In that case, $p({\rm det}|\theta)$ can be factorized out of the integrals, and acts as a normalization constant independent of $\Lambda$. This approximation is, however, not expected to hold for current data, due to large measurement errors.}
\end{enumerate}
It is important to note that our likelihood expressed in terms of the observable population, Eq.~\eqref{eq:convert}, is obtained through a simple rearrangement of terms in the standard likelihood, Eq.~\eqref{eq:std_diff_rate}, and is therefore equally valid.

We emphasize that even in this framework, selection effects must be incorporated into the hierarchical inference via the detection probability $p({\rm det}|\theta)$. They cannot be neglected during inference and applied only later, e.g., through Eq.~\eqref{eq:obs_rate}, to recover the astrophysical population. A potential drawback is that small values of $p({\rm det}|\theta)$ in the denominator may cause numerical instabilities. We discuss how we mitigate this in practice in the next subsection. On the other hand, $p({\rm det}|\theta)$ needs to be evaluated only for the finite set of posterior samples of the individual events.
In practice, the population likelihood is computed via Monte Carlo integration by replacing the individual-event likelihoods with their posteriors using Bayes' theorem. For example, Eq.~\eqref{eq:convert} becomes
\begin{align}
p(\{d\}|\Lambda) \propto e^{-N_{\rm O}}\prod_{i=1}^{N_\mathrm{E}}\frac{1}{N_{\mathrm{s},i}} \sum_{\theta_{i,j}\sim p(\theta_i|d_i)} \frac{{\rm d}N_{\rm O}/{\rm d}\theta_{i,j}(\Lambda)}{p({\rm det}|\theta_{i,j})\pi_{\rm PE}(\theta_{i,j})},\label{eq:obs_ppop_mc}
\end{align}
where $\pi_{\rm PE}(\theta)$ is the prior used in parameter estimation and $N_{\mathrm{s},i}$ is the number of posterior samples for event~$i$. The detection probabilities $p({\rm det}|\theta_{i,j})$ can thus be precomputed and stored.

In Appendices~\ref{app:marg_rate} and~\ref{app:marg_subset}, we describe how to marginalize over the number of events and over a subset of parameters. In Appendix~\ref{app:gaussian}, we illustrate that our approach correctly infers the observable population, contrasting with
%the flawed method presented
the previous approach pointed out to be flawed
in Ref.~\cite{Essick:2023upv}, using the same toy model as in that work.

\subsection{Application to O3 events}\label{sec:model}

The parameters that we model are
$\theta=(m_1,q,z,\chi_1,\chi_2,\cos\theta_1,\cos\theta_2)$, where $m_1$ is the primary source-frame mass, $q=m_2/m_1\leq 1$ is the binary mass ratio, $\chi_{1,2}$ are the dimensionless spin magnitudes, and $\theta_{1,2}$ are the angles between the spins and the orbital angular momentum.
We model the observable differential number of events as
\begin{align}
     \frac{{\rm d}N_{\rm O}}{{\rm d}\theta}(\Lambda) &= \frac{{\rm d}N_{\rm O}}{{\rm d}m_1{\rm d}z} (\Lambda)p_{\rm O}(q|m_1,\Lambda) \nonumber  \\ &\phantom{=}\ \times p_{\rm O}(\chi_1|\Lambda)p(\chi_2|\Lambda)p_{\rm O}(\cos \theta_1,\cos\theta_2|\Lambda). 
\end{align}
For the mass ratio, the spin magnitudes, and tilt angles, we assume the same functional forms as in the default model of Ref.~\cite{KAGRA:2021duu}, since those parameters play a smaller role in the detectability of a source than the primary mass and the redshift. 
Finally, we assume
\begin{align}
&\frac{{\rm d}N_{\rm O}}{{\rm d}m_1{\rm d}z}(\Lambda) =
	\begin{cases}
		\sum_{i=1}^{n_c} \lambda_iG(m_1|\mu_i,\sigma_i)\Gamma(z|\alpha_i,\theta_i) \nonumber \\ \quad {\rm if} \;  m_{\rm min}\leq m_1 \leq 100M_{\odot}  ;  \nonumber\\ 
		0 \ {\rm otherwise} ,
	\end{cases} \label{eq:m1_pdf} \\
\end{align}
where $G(m_1|\mu,\sigma)$ is a Gaussian distribution on $m_1$, with mean $\mu$ and standard deviation $\sigma$, $\Gamma(z|\alpha,\theta)$ is a gamma distribution on $z$ with shape and scale parameters $\alpha$ and $\theta$, and $\lambda_i$ are the weights of the components in the sum. The above expression is normalized to $N_{\rm O}$, so the weights do not sum to unity. This model differs from the \textsc{Power Law + Peak} model used by the LVK: it includes neither a power-law nor a smoothing function, and the number of Gaussians $n_c$ is allowed to vary and is inferred using the reversible-jump Markov Chain Monte Carlo (RJMCMC) implementation of the \texttt{Eryn} sampler~\cite{Karnesis:2023ras}. Our model allows for correlations between $m_1$ and $z$, since each Gaussian component in $m_1$ is associated with its own gamma distribution in $z$. Such a correlation is naturally expected to arise in the observable space, because the detectability of a signal depends strongly on both the primary mass and the redshift.
In full generality, correlations involving the remaining parameters, $q$, $\chi_{1,2}$, and $\cos\theta_{1,2}$, are expected to arise already in the astrophysical space, as a consequence of the different formation channels of compact binaries. Since the detectability of a signal depends more weakly on these parameters than on $m_1$ and $z$, such correlations are not expected to be significantly enhanced in the observable space. We therefore retain a separable form for $q$, $\chi_{1,2}$, and $\cos\theta_{1,2}$, in line with the fiducial model of Ref.~\cite{KAGRA:2021duu}.

\begin{figure}
\includegraphics[width=0.98\columnwidth]{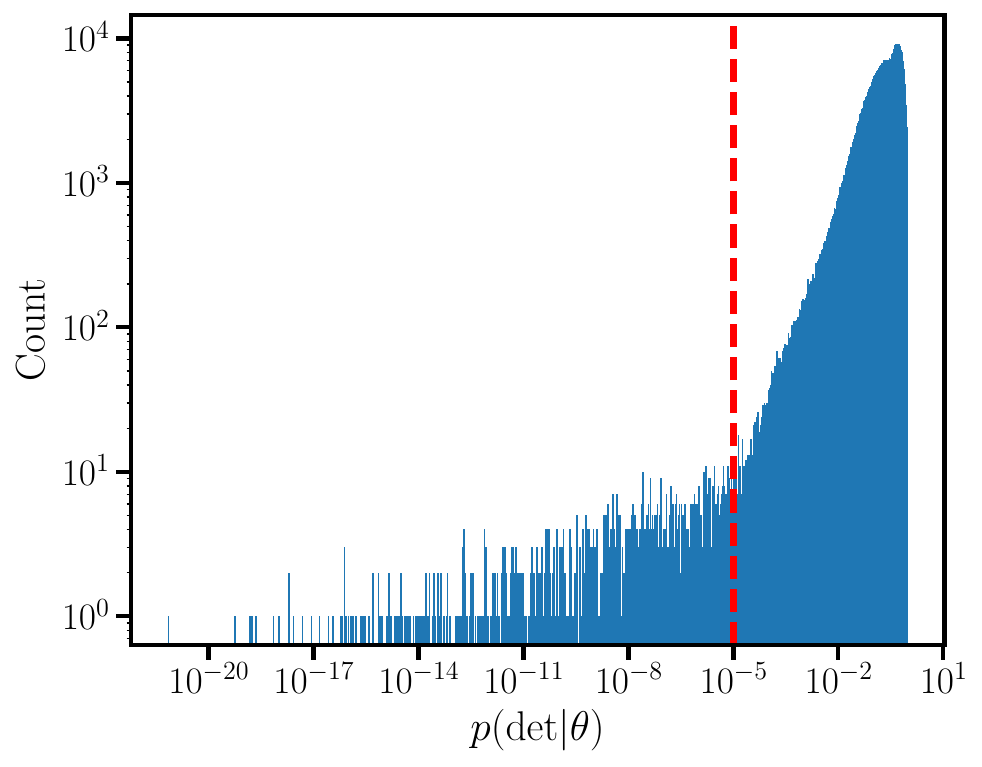}
    \caption{Distribution of detection probabilities estimates on the posterior samples of the 59 events included in the analysis. The red dashed line at $10^{-5}$ marks a transition in the distribution, interpreted as the point below which the estimates are dominated by numerical noise.
}
\label{fig:pdets}
\end{figure}

 We use the emulator specific to the third LVK observing run (O3)
from Ref.~\cite{Callister:2024qyq} to compute $p({\rm det}|\theta)$ on the public posterior samples for the 59 O3 events\footnote{We use the same samples as the LVK for their O3-only analysis with the \textsc{Power Law + Peak} model~\cite{ligo_scientific_collaboration_2024_11254021}.}. %We thus restrict to the 59 O3 events, excluding the 10 earlier detections included in Ref.~\cite{KAGRA:2021vkt}. This leads to minor differences from the full analysis. 
We plot in Fig.~\ref{fig:pdets} the values of $p({\rm det}|\theta)$ estimated on all the posterior samples used in the analysis. A transition in the distribution appears around $10^{-5}$, indicated by the red vertical line,
%which we interpret as
a limitation of the numerical accuracy of the emulator used for the interpolation. Values below $10^{-5}$ are dominated by numerical noise and should therefore be considered unreliable. 
Thus, we remove from our analysis all the posterior samples whose estimated $p({\rm det}|\theta)$ is below $10^{-5}$, in order to avoid numerical instabilities in the Monte Carlo integrals. This can be interpreted as enforcing $\dd{N_\mathrm{O}}/\dd{\theta}$ to vanish wherever $p({\rm det}|\theta)<10^{-5}$. 

As a rule of thumb, the accuracy required on $p({\rm det}|\theta)$ can be estimated through the requirement $\int_{\mathcal{V}} {\rm d} \theta \  p({\rm det}|\theta){{\rm d}N_{\rm A}}/{{\rm d}\theta} \leq 1$, which means requiring that the differential astrophysical number of events and the detection probability are such that we expect to have detected less than one event with parameters in the volume $\mathcal{V}$ during the observation time. We can get a lower limit on the required accuracy by estimating the astrophysical number of events within the observable region of parameter space:
\begin{equation}
    N_{\mathrm{A},{\rm obs}}=\int {\rm d} \theta\frac{1}{p({\rm det}|\theta)}\frac{{\rm d}N_{\rm O}}{{\rm d}\theta}.  
\end{equation}
Treating the observed events as fair draws from the observable differential rate, we get 
\begin{equation}
    N_{\mathrm{A},{\rm obs}}=\sum_{i=1}^{N_\mathrm{E}} \frac{1}{p({\rm det}|\theta_i)}.  
\end{equation}
Finally, we estimate $p({\rm det}|\theta_i)$ as the mean of $p({\rm det}|\theta)$ over the posterior samples of each event. The resulting accuracy limit on $p({\rm det}|\theta)$ is therefore $\sim 1/N_{\mathrm{A},{\rm obs}}$. For the 59 events used in this analysis, this gives $p({\rm det}|\theta) \geq 2\times10^{-3}$. Therefore, this
% rapid
back-of-the-envelope
estimate suggests that cutting samples with detection probability below $10^{-5}$ should not significantly affect the results. 
For future analyses, pushing the numerical accuracy of such selection-effect emulators is a key area for improvement. As a test, we verified that the results we present here remain unchanged if instead we cut samples with detection probability below $10^{-3}$.   

We emphasize that the emulator used to compute $p({\rm det}|\theta)$ was trained over the injections used to estimate the selection function. This multidimensional parameter space is likely broader than the region spanned by the posterior samples, and in this sense, the training is not optimal.
Since detection probabilities only need to be evaluated at the locations of the posterior samples, the training dataset could instead be tailored to focus on this region. Such targeted training would allow the emulator to learn the detection probabilities more accurately where it matters the most.

\begin{figure*}
    \centering

\includegraphics[height=0.3\textwidth]{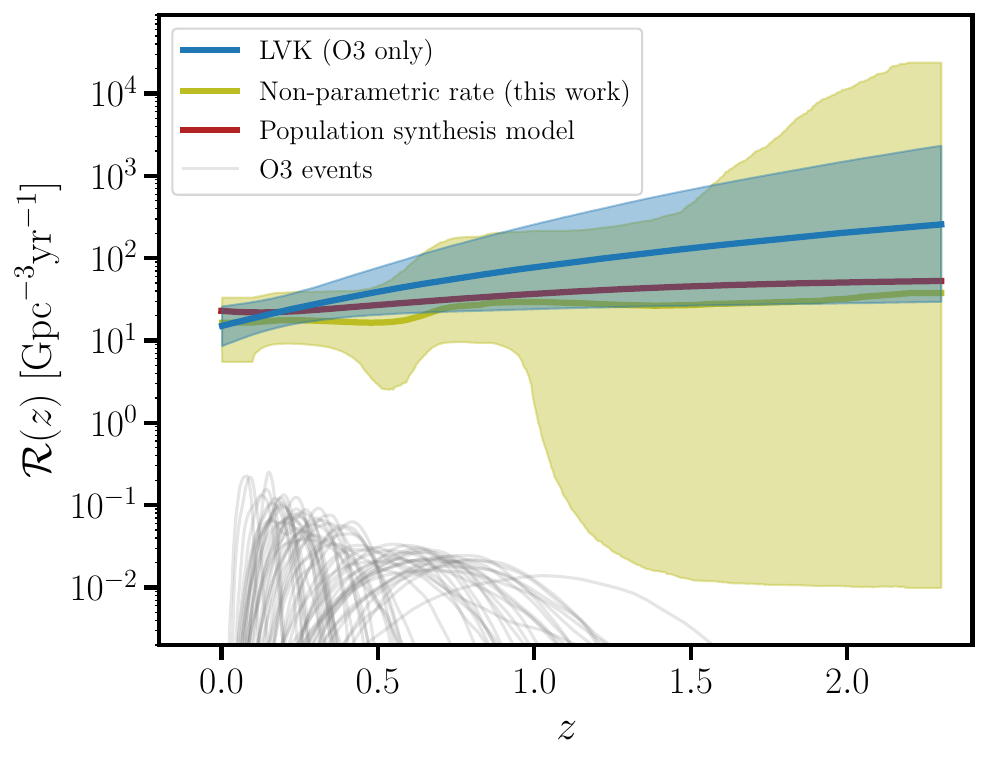}
\includegraphics[height=0.3\textwidth]{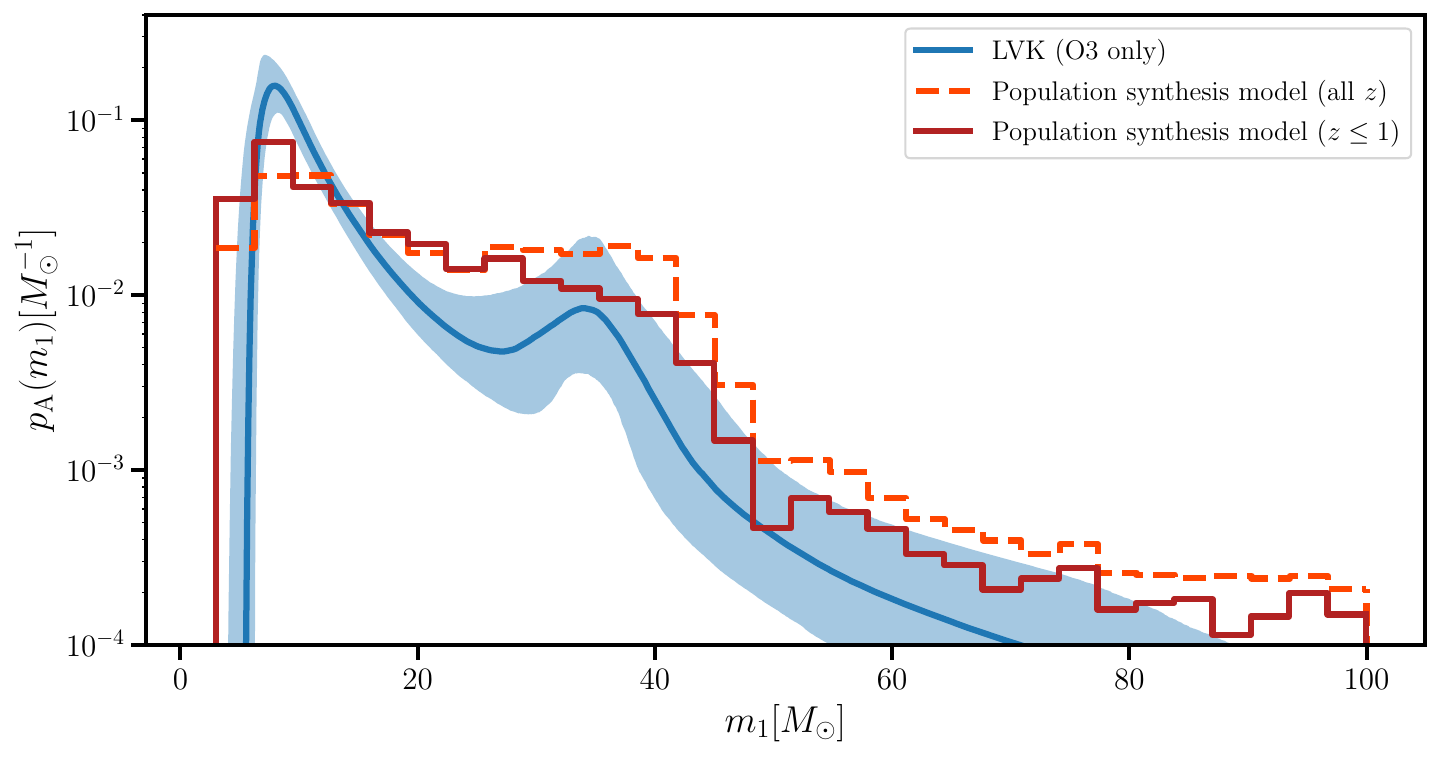}\label{fig:comp_astro_lvk}
\includegraphics[height=0.3\textwidth]{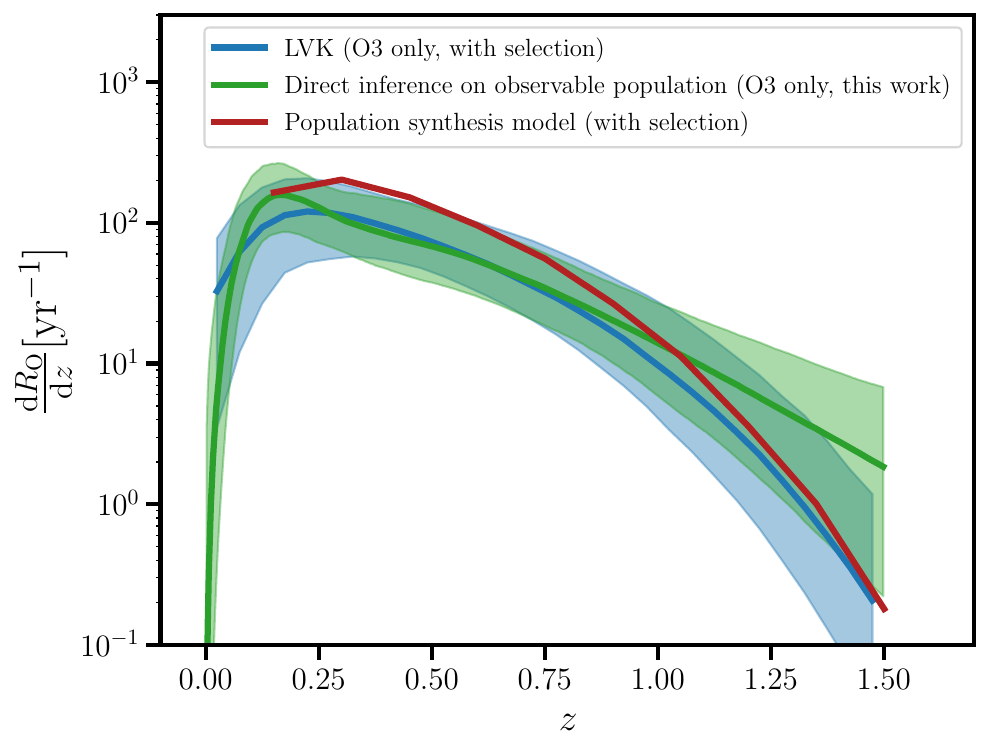}
\includegraphics[height=0.3\textwidth]{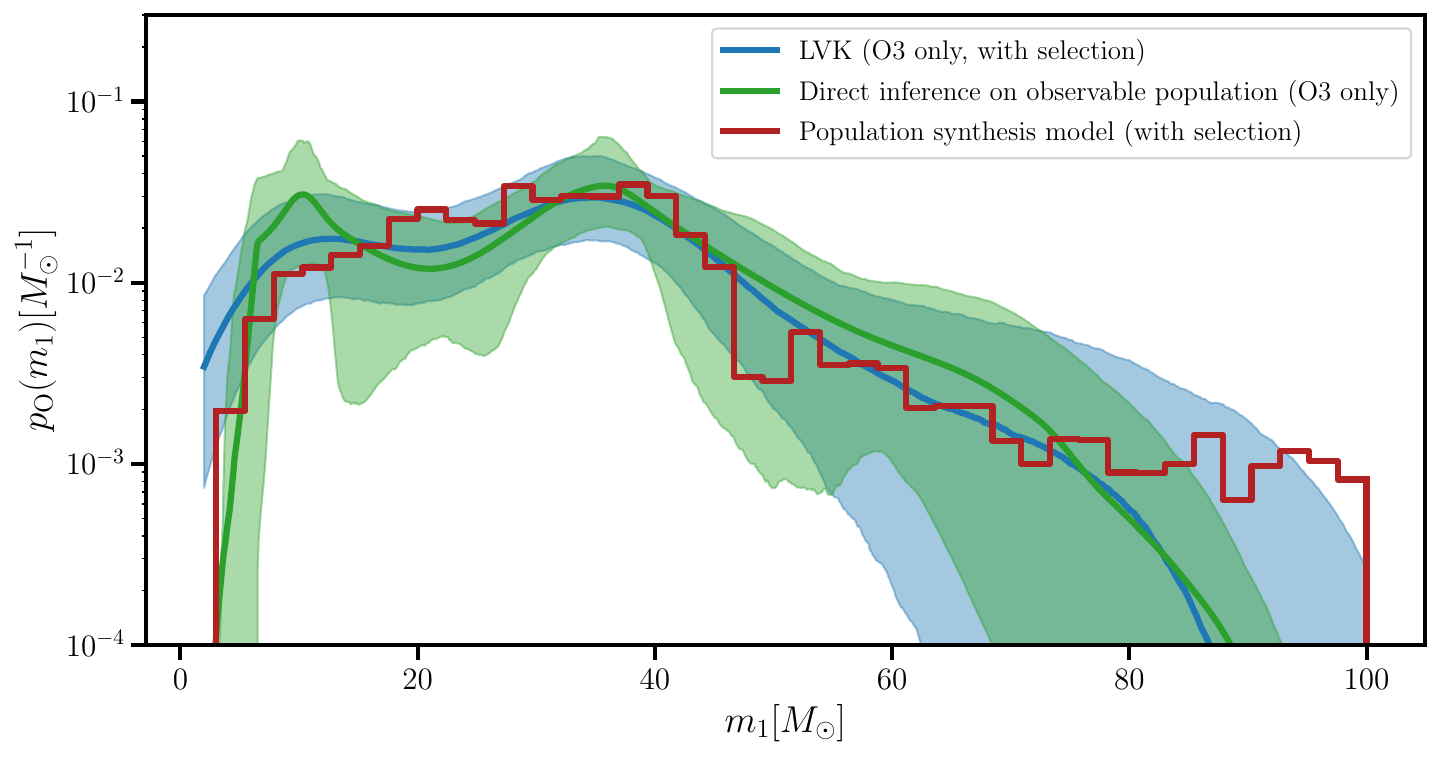}
    
    \caption{Comparison between the inferred population and the population-synthesis model of Ref.~\cite{2023MNRAS.520.5259A}. The top row shows results in the astrophysical space, whereas the bottom row shows the comparison in the observable space. In all plots, shaded areas show the $90\%$ credible region and solid lines inside show the medians.
    {\it Top left}: Evolution of the volumetric rate with redshift using a power-law model (blue) and a non-parametric reconstruction (yellow). The gray curves at the bottom of the plot are the marginalized redshift posterior distributions of the O3 events (the heights of these curves are arbitrary).  {\it Top right}: Astrophysical mass distribution inferred by the LVK compared to the predictions of the population-synthesis model with all mergers (orange) and only up to $z\leq 1$ (red). {\it Bottom left}: Rate of observable mergers as a function of redshift from our direct inference (green), from the LVK results reweighted by the detection probability (blue), and from the population-synthesis model similarly reweighted (red). {\it Bottom right}: Direct inference of the observable mass distribution (green) is compared to the LVK inference on the astrophysical distribution reweighted by the detection probability (blue), and to the population-synthesis model to which we apply this same procedure (red). 
    In the observable space, the population-synthesis model shows even better agreement with the inference on O3 events than
    when the astrophysical population is restricted to $z \leq 1$, except for the high-mass end. }\label{fig:plots}
\end{figure*}

\section{Comparing data and models}\label{sec:pop_astro}

We now illustrate the advantages of comparing population-synthesis models directly in the observable space rather than in the astrophysical space. In Fig.~\ref{fig:comp_astro_lvk}, we compare the predictions of the population-synthesis model of Ref.~\cite{2023MNRAS.520.5259A}, described in Appendix~\ref{app:pop_synth}, with both the inferred astrophysical population (top row) and the inferred observable distribution (bottom row) from O3. The model used for the observable population is described in Sec.~\ref{sec:model}. For the astrophysical population, we use the fiducial LVK model~\cite{ligo_scientific_collaboration_2024_11254021}. We emphasize that our goal here is to highlight the benefits of this approach, and our discussion is not tied to the specific population-synthesis model used for illustration.

First, we show the inferred estimate of the volumetric rate in the top-left panel of Fig.~\ref{fig:plots}, together with redshift posteriors of the individual events.
The LVK analysis assumes that the astrophysical volumetric rate of BBH mergers is given by~\cite{Fishbach:2018edt,KAGRA:2021duu} 
\begin{align}
    \mathcal{R}(z) &= \mathcal{R}_0(1+z)^{\kappa},\label{eq:pl_z}
\end{align}
We also present the results obtained using a more flexible (non-parametric) model for $\mathcal{R}(z)$. More specifically, we model $\log_{10}\mathcal{R}(z)$ as a series of step functions, where both the values of the function and the number and positions of the knots are allowed to vary freely, which is achieved via RJMCMC. The flexible reconstruction reveals significantly larger uncertainties at $z \gtrsim 1$, reflecting the limited number of observed events in this region. As a result, any inference beyond $z \sim 1$ with the fiducial LVK model in Eq.~\eqref{eq:pl_z} relies on extrapolation. The prediction from our fiducial population-synthesis model is within the central 90\% uncertainties of both reconstruction methods in the considered redshift range, indicating there is no evidence to rule out that prediction based on this dataset. We stress that the lower bound of our non-parametric reconstruction is set by the prior on $\log_{10}\mathcal{R}(z)$. Modeling the logarithm of the volumetric rate allows us to capture variations across several orders of magnitude, but it also prevents $\mathcal{R}(z)$ from reaching zero. In regions where the data are uninformative, the reconstruction is therefore governed by the interplay between the imposed prior and the selection function.

Next, we turn to the astrophysical distribution of BBH masses. The fiducial LVK analysis~\cite{KAGRA:2021duu} assumes that the primary-mass distribution is described by the sum of a power-law component and a Gaussian component, modulated by a tapering function that ensures a smooth falloff at low masses, referred to as the \textsc{Power Law + Peak} model \cite{Talbot:2018cva}. Crucially, the mass distribution is assumed to be the same across all redshifts. However, as discussed previously in the context of the merger rate, the range of validity of the current results appears to extend up to $z\!\sim\!1$, beyond which observational constraints weaken due to the scarcity of detected events.

In the top-right panel of Fig.~\ref{fig:plots}, we illustrate the importance of taking this into account when comparing astrophysical models to the inferred distributions. We show the probability density function of BBH as a function of the primary mass ($m_1$) inferred with the \textsc{Power Law + Peak} model, together with the predictions of the population-synthesis model. When including all the mergers in the simulation, the prominence of the first peak is underestimated, while the second peak and the high-mass end of the distribution are overestimated. On the other hand, when restricting to mergers at $z\leq1$, the predictions of the model are in better agreement with the LVK analysis, although some discrepancy remains. This suggests that comparing conditional rather than overall astrophysical distributions is more robust \cite{Callister:2023tgi}, but requires knowing which regions of parameter space are suitable for comparison in the first place.

 Let us now turn to the comparison in the observable space. Using the methods outlined below, we infer directly the observable population. The bottom-left panel of Fig.~\ref{fig:plots} compares the predicted observational rate as a function of redshift from our inference, the LVK result, and the population-synthesis model. All three are in good agreement. 

The bottom-right panel of Fig.~\ref{fig:plots} shows the inferred distribution of observable events as a function of primary mass, compared with the LVK inference from O3 \cite{KAGRA:2021duu} and the population-synthesis model, both weighted by the detection probability. The LVK result is broadly consistent with our inference. The population-synthesis model also more closely matches the data once selection effects are accounted for, up to $\sim 80,M_{\odot}$, beyond which it appears to overpredict the abundance of massive systems.
These results are consistent with our manual restriction of the comparison to the region where the data are informative, achieved by selecting mergers at $z \leq 1$, as shown in Fig.~\ref{fig:plots}. While a naive comparison between the population-synthesis model and the inferred astrophysical distributions would suggest
% an excess of systems around $35 M_{\odot}$ and a deficit around $10 M_{\odot}$,
the former overpredicts the number of mergers around $\sim25M_\odot$ and $\sim40M_\odot$ and underpredicts the number at $\sim10M_\odot$,
the comparison in the observable space reveals this is not the case.

This confirms the value of performing comparisons directly on the observable populations, as the domain of validity of the models used in the inference is naturally taken into account. Performing such a comparison in a statistically robust manner, without resorting to the two-step procedure of deconvolving selection effects only to fold them back in, is precisely what is enabled by the approach we proposed in Sec.~\ref{sec:obs_pop}. We stress that inferring the joint $m_1$--$z$ distribution would mitigate part of the problem, but it would not solve the issues of extrapolation and the limited range of validity that arise when using parametric models. A more complex model for the redshift distribution could account for this, for example by allowing a transition with different power-law exponents on each side, and obtaining an uninformative posterior on the high-redshift exponent. However, this would purposefully introduce parameters that are meant to remain unconstrained.

\section{Conclusions}\label{sec:ccl}
Comparing populations inferred using parametric or non-parametric methods to astrophysical models requires a careful understanding of the domain over which the inferences are valid. Parametric models extrapolate constraints based on their assumed functional forms beyond the region of parameter space where observational data lie and the model is effectively constrained. Increasing the complexity of the models might cure this behavior, but at the cost of introducing additional parameters that are expected to be poorly constrained. Ultimately, non-parametric models can indicate where information is actually present by exhibiting increasingly large uncertainties in regions with little or no data. However, their behavior in such regions is heavily influenced by the choice of prior, which can complicate both the inference and the interpretation of the results—especially in multi-dimensional reconstructions.
As the number of GW observations increases, we require more accurate inference and means of comparison in order to make population analyses a precision science and maximize the ability of data to inform us on the formation of compact binaries.

A natural solution to this issue is to perform the comparison in the observable space. In this paper, we have demonstrated how to perform unbiased inference directly in the observable space and highlighted the value of this approach by comparing the results of a fiducial population-synthesis model in both the astrophysical and observable spaces. We used the population-synthesis model of Refs.~\cite{2019MNRAS.482.2991A, 2020ApJ...894..133A, 2023MNRAS.520.5259A, Arca_sedda_prep} to illustrate the questions addressed in this work; however, the central findings are not dependent on this specific model.

Our inference on the observable population is in remarkable agreement with the LVK results after reweighing the latter by the selection function. When comparing with the population-synthesis model, we find that performing the comparison in the observable space confirms the intuition acquired by restricting the astrophysical population to mergers which are well within the detector horizon.

An important aspect of inferring the observable distribution is that selection effects still need to be properly accounted for. Unlike inference on the astrophysical population—where the likelihood involves integrating the detection probability over the population—here the likelihood depends directly on the detection probability for individual signals, marginalized over possible data realizations. As a result, all necessary values can be precomputed from the posterior samples and reused during the population inference. However, since these values enter the denominator of the likelihood, numerical inaccuracies can compromize the inference. To mitigate this, it is essential to assess both the accuracy of the method used to estimate these quantities and the level of precision required. 

In this work, we used the emulator developed in Ref.~\cite{Callister:2024qyq} to compute detection probabilities and verified that its current accuracy is sufficient for our purposes. It is important to note that the set of search injections on which the emulator was trained was originally designed for inferring the astrophysical population. Further improvements could be achieved by training the emulator on an injection set specifically optimized for inference on the observable population. Otherwise, the use of physically motivated approximants~\cite{Essick:2023toz,Gerosa:2024isl,Lorenzo-Medina:2024opt} may mitigate this issue.
Understanding the impact of Monte Carlo estimator accuracy and how it scales with the number of events in the direct inference of the observable population—analogous to the investigations in Refs.~\cite{Talbot:2023pex,Essick:2022ojx,Farr:2019rap, Heinzel:2025ogf} for the standard approach—will be important. We leave a detailed investigation of this to future work.

In principle, the astrophysical distribution can be recovered from the observable one by dividing by the detection probability. However, we find this approach to be prone to numerical instabilities in regions where the detection probability is low. These are the same regions in which constraints from both parametric and nonparametric models fit in the astrophysical space are not to be trusted anyway. This does not invalidate the inference of the observable population, as the observed events lie precisely in the portion of parameter space where the detection probability is high (within the accuracy of the emulator, as discussed above).
%}  

In this work, we limited ourselves to the O3 data set, as the tools required for our analysis, most importantly an estimator of $p({\rm det}|\theta)$, were available only for these data. We are currently computing the corresponding ingredients for the newly released data from the first part of the fourth LVK observing run~\cite{LIGOScientific:2025slb}, which will allow us to extend our study and to more robustly quantify the agreement between the observations and population-synthesis predictions.
For instance, a fully non-parametric reconstruction could be carried out directly on the observable population. Because the observable distribution vanishes outside a well-defined region of parameter space, inference in this domain is intrinsically easier: unlike for the astrophysical population, non-parametric methods do not need to recover the prior in regions where the data are uninformative. Approaches such as those described in Refs.~\cite{Fabbri:2025faf,Rinaldi:2025evs,Alvarez-Lopez:2025ltt} could then be used to map these reconstructions onto parametric or astrophysically motivated models, enabling a principled assessment of their goodness of fit.
%}

The main takeaway of this paper is that of Fig.~\ref{fig:plots}: while our fiducial population-synthesis model appears to be in some disagreement with the inferred population
in the astrophysical space, it lies well within the statistical uncertainties of our measurements over most of the mass spectrum when compared in the observable space. A drawback of performing the comparison in observable space is that the resulting metrics, such as reconstructed correlations, can be less straightforward to interpret directly in astrophysical terms. However, a model agrees with data when it can well predict observables; for this reason, confronting theoretical models with measurements in observable space provides a valuable alternative approach to assess their consistency with the data.

\section*{Acknowledgments}
We thank 
Reed Essick, 
Tassos Fragos, 
Jack Heinzel,
Daniel Holz,
Colm Talbot,
and Salvatore Vitale
for discussions, and Ana Lorenzo-Medina for helping us catch an error in the computation of the detection probabilities. We thank the organizers and participants of the {\it ``Gravitational-wave snowballs, populations, and models''} (Sexten Center for Astrophysics) and the {\it ``Emerging methods in GW population inference''} (IFPU, Trieste) workshops, where
some of these ideas were first discussed.
A.T., D.G., T.B., R.B. and R.T. are supported by MUR Grant ``Progetto Dipartimenti di Eccellenza 2023-2027'' (BiCoQ) and the ICSC National Research Centre funded by NextGenerationEU. A.T., D.G., T.B. and R.T. are supported by ERC Starting Grant No.~945155--GWmining, Cariplo Foundation Grant No.~2021-0555, MUR PRIN Grant No.~2022-Z9X4XS, and Italian-French University (UIF/UFI) Grant No.~2025-C3-386. A.T. and D.G. are supported by MUR Young Researchers Grant No. SOE2024-0000125.
D.G. is  supported by MSCA Fellowship No.~101064542--StochRewind and MSCA Fellowship No.~101149270--ProtoBH. R.B. is  supported by the Italian Space Agency grant Phase B2/C activity for the LISA mission, agreement No. 2024-NAZ-0102/PER.
M.M. is supported by the Royal Commission for the Exhibition of 1851 Research Fellowship and by the LIGO Laboratory through the National Science Foundation awards No. PHY-1764464 and No. PHY-2309200.
S.R. is funded by Deut\-sche For\-schungs\-ge\-mein\-schaft (DFG, German Research Foundation) project No. 546677095.
This material is based upon work supported by NSF's LIGO Laboratory which is a major facility fully funded by the National Science Foundation, and has made use of data or software obtained from the Gravitational Wave Open Science Center (\href{https://gwosc.org}{gwosc.org}), a service of the LIGO Scientific Collaboration, the Virgo Collaboration, and KAGRA.

\appendix

\section{Marginalization over the rate}\label{app:marg_rate}
Treating $N_{\rm A}$ as an independent parameter, we have
\begin{align}
p(\{d\}|\Lambda,\!N_{\rm A}) \!\propto\! N_{\rm A}^{N_\mathrm{E}} e^{-N_{\rm A}p({\rm det}|\Lambda)}\prod_i^{N_\mathrm{E}}\!\int \!\!{\rm d}\theta_i   p(d_i|\theta_i)p_{\rm A}(\theta_i|\Lambda) .\label{eq:std_rate}
\end{align}
Notice that the terms depending on $N_{\rm A}$ are completely independent of those depending on $\Lambda$. Therefore, if we assume a separable prior, $\pi(\Lambda,N_{\rm A}) = \pi(\Lambda) \pi(N_{\rm A})$ \cite{Essick:2023upv}, we have 
\begin{equation}
   p(\{d\}|\Lambda) \propto p({\rm det}|\Lambda)^{-N_\mathrm{E}}\prod_i^{N_\mathrm{E}}\int {\rm d}\theta_i \  p(d_i|\theta_i)p_{\rm A}(\theta_i|\Lambda). \label{eq:std}   
\end{equation}
We can follow a similar approach for the observable population. 
Treating $N_{\rm O}$ as an independent variable, we can write 
\begin{align}
p(\{d\}|\Lambda,N_{\rm O}) \propto N_{\rm O}^{N_\mathrm{E}}  e^{-N_{\rm O}}\prod_i^{N_\mathrm{E}}\int {\rm d}\theta_i \frac{p(d_i|\theta_i)}{p({\rm det}|\theta_i)}p_{\rm O}(\theta_i|\Lambda).\label{eq:convert_pobs}
\end{align}
Assuming a separable prior $\pi(\Lambda,N_{\rm O}) = \pi(\Lambda) \pi(N_{\rm O})$, marginalization over the observable number of events yields\footnote{As pointed out in Ref.~\cite{Essick:2023upv} only a prior $\propto N_{\rm O}^{-1}$ results in any case in a separable prior also for the astrophysical rate, which is then $\propto N_{\rm A}^{-1}$.}
\begin{align}
p(\{d\}|\Lambda) \propto \prod_i^{N_\mathrm{E}}\int \dd{\theta_i} \frac{p(d_i|\theta_i)}{p({\rm det}|\theta_i)} p_{\rm O}(\theta_i|\Lambda).
\label{eq:likelihood_pobs}
\end{align}

\section{Marginalization over a subset of parameters}\label{app:marg_subset}
In many situations we are interested in only a subset of parameters, and want to marginalize over the remaining ones. Let us split the set of parameters as $\theta=(\vartheta,\bar\vartheta)$. $\vartheta$ typically stands for the masses, spins and redshift, while $\bar\vartheta$ stands for sky location, polarisation, phase and inclination. Using Bayes' theorem, we can write Eq.~\eqref{eq:std_rate} as
\begin{align}
    &p(\{d\}|\Lambda,N_{\rm A}) \propto N_{\rm A}^{N_\mathrm{E}} e^{-N_{\rm A}p({\rm det}|\Lambda)} \notag \\ &\times \prod_i^{N_\mathrm{E}}\int {\rm d}\vartheta_{i} {\rm d}{\bar\vartheta}_{i} \frac{p(\vartheta_{i},\bar\vartheta_{i}|d_i)}{\pi_{{\rm PE},i}(\vartheta_{i},\bar\vartheta_{i})})p_{\rm A}(\vartheta_{i},\bar\vartheta_{i}|\Lambda),\label{eq:std_sep1} 
\end{align}
Assuming that the population and the parameter estimation priors are separable in $\vartheta$ and $\bar\vartheta$, that $\pi_{{PE},i}(\bar\vartheta)$ is the same for all events, and that $p_{A}(\bar\vartheta|\Lambda)=\pi_{\rm PE}(\bar\vartheta)$,\footnote{These assumptions are commonly made in LVK analyses.} we get
\begin{align}
    p(\{d\}|&\Lambda, N_{\rm A}) \propto N_{\rm A}^{N_\mathrm{E}} e^{-N_{\rm A}p({\rm det}|\Lambda)} \nonumber \\ &\times \prod_i^{N_\mathrm{E}}\int {\rm d}\vartheta_{i} {\rm d}{\bar\vartheta}_{i} \frac{p(\vartheta_{i},\bar\vartheta_{i}|d_i)}{\pi_{{\rm PE},i}(\vartheta_{i})})p(\vartheta_{i}|\Lambda)\label{eq:std_sep2}. 
\end{align}
The marginalization over $\bar\vartheta$ can be readily performed to obtain
\begin{equation}
    p(\{d\}|\Lambda,N_{\rm A}) \propto \! N_{\rm A}^{N_\mathrm{E}} e^{-N_{\rm A}p({\rm det}|\Lambda)} \prod_i^N \!\int \!\! {\rm d}\vartheta_{i}  \frac{p(\vartheta_{i}|d_i)}{\pi_{{\rm PE},i}(\vartheta_{i})})p(\vartheta_{i}|\Lambda)\label{eq:std_sep3}. 
\end{equation}
Next, we define the detection probability marginalized over $\bar\vartheta$ as
\begin{equation}
    p({\rm det}|\vartheta)=\int  {\rm d} \bar\vartheta \ p_{\rm A}({\rm det}|\vartheta,\bar\vartheta)\pi_{\rm PE}(\bar\vartheta).
\end{equation}
Performing, the same trick that lead to Eq.~\eqref{eq:convert}, but using $p({\rm det}|\vartheta)$, we get
\begin{equation}
     p(\{d\}|\Lambda,N_{\rm O}) \propto N_{\rm O}^{N_\mathrm{E}} e^{-N_{\rm O}} \prod_i^{N_\mathrm{E}}\int {\rm d}\vartheta_{i} \frac{p(d_i|\vartheta_{i})}{p({\rm det}|\vartheta_{i})}p_{O}(\vartheta_{i}|\Lambda)\label{eq:right_inf_obs_marg}\,.
\end{equation}
This is similar to Eq.~\eqref{eq:convert_pobs}, but with the probability of detection being marginalized over $\bar\vartheta$.

\section{Gaussian toy model}\label{app:gaussian}
To further illustrate that Eq.~\eqref{eq:convert} provides an unbiased inference, we rework the Gaussian toy model presented in Ref.~\cite{Essick:2023upv}. In this toy model, the hyperparameters are $\lambda = ({\mu_\Lambda}, \sigma_\Lambda)$ and one assumes
\begin{align}
    p_{\rm A}(\theta|\Lambda) &= (2\pi \sigma_\Lambda^2)^{-1/2} \exp\left[ -\frac{(\theta-\mu_\Lambda)^2}{2\sigma_\Lambda^2}\right]\,, \\
    p(d|\theta) &= (2\pi \sigma_e^2)^{-1/2} \exp\left[ -\frac{(d-\theta)^2}{2\sigma_e^2}\right]\,, \\
    p({\rm det}|d) &= \exp\left[ -\frac{(d-\mu_D)^2}{2\sigma_D^2}\right]\,,
\end{align}
such that 
\begin{align}
    p(d|\Lambda) &= \left[2\pi (\sigma_\Lambda^2 + \sigma_e^2) \right]^{-1/2} \exp\left[ -\frac{(d-\mu_\Lambda)^2}{2(\sigma_\Lambda^2 + \sigma_e^2)} \right]\,,\\
    p({\rm det}|\Lambda) &= \left( \frac{\sigma_D^2}{\sigma_\Lambda^2 + \sigma_e^2 + \sigma_D^2} \right) \exp\left[ -\frac{(\mu_D - \mu_\Lambda)^2}{2(\sigma_\Lambda^2 + \sigma_e^2 + \sigma_D^2)} \right]
    .
\end{align}

%\prlsubsec{Fitting the intrinsic distribution}
 Assuming a flat prior on $\Lambda$ and using Eq.~\eqref{eq:std} to obtain the posterior returns
  \cite{Essick:2023upv}
\begin{align}
    p(\Lambda| \{d\}) &\propto
      \left[\frac{\sigma_\Lambda^2 + \sigma_e^2 + \sigma_D^2}{2\pi \sigma_D^2 (\sigma_\Lambda^2 + \sigma_e^2)}\right]^{N_\mathrm{E}/2} \nonumber \\
      &\times \exp\left[ -\frac{\sum_i^{N_\mathrm{E}} (d_i-\mu_\Lambda)^2}{2(\sigma_\Lambda^2 + \sigma_e^2)} + \frac{N_\mathrm{E}(\mu_D - \mu_\Lambda)^2}{2(\sigma_\Lambda^2 + \sigma_e^2 + \sigma_D^2)} \right]\,. \label{eq:i like this eq}
\end{align}

Reference~\cite{Essick:2023upv} then defines point estimators by maximizing Eq.~\eqref{eq:i like this eq} with respect to $\mu_\Lambda$ and $\sigma_\Lambda^2$:
\begin{align}
   \hat{\mu}_\Lambda & = \frac{(\sigma_\Lambda^2 + \sigma_e^2 + \sigma_D^2) m_d - (\sigma_\Lambda^2 + \sigma_e^2) \mu_D}{\sigma_D^2} \label{eq:hat mu lambda}\,, \\
   \hat{\sigma}^2_\Lambda & = \frac{\sigma_D^2 (m_{d^2} - m_d^2)}{\sigma_d^2 - (m_{d^2} - m_d^2)} - \sigma_e^2\,, \label{eq:hat sigma lambda}
\end{align}
where
\begin{align}
    m_d & = \frac{1}{N_\mathrm{E}}\sum\limits_i^{N_\mathrm{E}} d_i \,,\\
    m_{d^2} & = \frac{1}{N_\mathrm{E}}\sum\limits_i^{N_\mathrm{E}} d_i^2\,,
\end{align}
are the moments of the observed data.
The event and data distributions conditioned on detection are given by
\begin{align}
   & p(\theta|{\rm det},\Lambda)  = \left[ \frac{\sigma_\Lambda^2 + \sigma_e^2 + \sigma_D^2}{2\pi \sigma_\Lambda^2 (\sigma_e^2 + \sigma_D^2)} \right]^{1/2} \nonumber \\ 
    &\times \exp\left\{ -\frac{\sigma_\Lambda^2 + \sigma_e^2 + \sigma_D^2}{2\sigma_\Lambda^2 (\sigma_e^2 + \sigma_D^2)} \left[ \theta - \frac{\sigma_\Lambda^2 \mu_D + (\sigma_e^2 + \sigma_D^2)\mu_\Lambda}{\sigma_\Lambda^2 + \sigma_e^2 + \sigma_D^2} \right]^2 \right\} \label{eq:gaussian3 good theta given det} \\
    &p(d|{\rm det},\Lambda)  = \left[ \frac{\sigma_\Lambda^2 + \sigma_e^2 + \sigma_D^2}{2\pi(\sigma_\Lambda^2 + \sigma_e^2)\sigma_D^2}\right]^{1/2} \nonumber \\ &\times \exp\left\{ -\frac{\sigma_\Lambda^2 + \sigma_e^2 + \sigma_D^2}{2(\sigma_\Lambda^2 + \sigma_e^2)\sigma_D^2} \left[ d - \frac{\sigma_D^2 \mu_\Lambda + (\sigma_\Lambda^2 + \sigma_e^2)\mu_D}{\sigma_\Lambda^2 + \sigma_e^2 + \sigma_D^2} \right]^2 \right\} \label{eq:gaussian3 good data given det}
\end{align}
With Eq.~\eqref{eq:gaussian3 good data given det} in hand, Ref.~\cite{Essick:2023upv} shows that the moments of the detected data will approach
\begin{align}
    & \lim\limits_{N_\mathrm{E}\rightarrow\infty} m_d  = \frac{\sigma_D^2 \mu_\Lambda + (\sigma_\Lambda^2 + \sigma_e^2)\mu_D}{\sigma_\Lambda^2 + \sigma_e^2 + \sigma_D^2} \,,\label{eq:gaussian2 data m1} \\
    & \lim\limits_{N_\mathrm{E}\rightarrow\infty} \left( m_{d^2} - m_d^2 \right)  = \frac{(\sigma_\Lambda^2 + \sigma_e^2)\sigma_D^2}{\sigma_\Lambda^2 + \sigma_e^2 + \sigma_D^2} \,.\label{eq:gaussian2 data m2}
\end{align}
Plugging Eqs.~\eqref{eq:gaussian2 data m1} and~\eqref{eq:gaussian2 data m2} into Eqs.~\eqref{eq:hat mu lambda} and~\eqref{eq:hat sigma lambda} yields
\begin{align}
    \lim\limits_{N_\mathrm{E}\rightarrow\infty} \hat{\mu}_\Lambda & = \mu_\Lambda \,,\\
    \lim\limits_{N_\mathrm{E}\rightarrow\infty} \hat{\sigma}^2_\Lambda & = \sigma^2_\Lambda \,,
\end{align}
which demonstrates that the inference is unbiased.

%\prlsubsec{Fitting the observed distribution: naive approach with a mistake}

Next, Ref.~\cite{Essick:2023upv} considers the approach of fitting the observed distribution, using Eq.~\eqref{eq:likelihood_pobs}, but without the $p({\rm det}|\theta)$ term in the denominator. Because all the distributions are Gaussian, they adopt a Gaussian ansatz for the observable distribution:
\begin{equation}
    p_{\rm O}(\theta|\Lambda) = (2\pi\sigma^2)^{-1/2} \exp\left[ -\frac{(\theta-\mu)^2}{2\sigma^2}\right].
\end{equation}
The posterior under this model is
%\begin{multline}
%    q(\Lambda|\{d_i,D_i\}, N) \propto \\ (2\pi (\sigma^2 + \sigma_e^2))^{-N/2} \exp\left( -\frac{\sum_i^N (d_i-\mu)^2}{2(\sigma^2 + \sigma_e^2)} \right)
%\end{multline}
\begin{align}
    p(\Lambda|\{d_i,{\rm det}_i\}, N_\mathrm{E}) \propto [2\pi (\sigma^2 + \sigma_e^2)]^{-N_\mathrm{E}/2} \nonumber \\  \times \exp\left[ -\frac{\sum_i^{N_\mathrm{E}} (d_i-\mu)^2}{2(\sigma^2 + \sigma_e^2)} \right],
\end{align}
and the associated estimators are
\begin{align}
    \hat{\mu} & = m_d \label{eq:gaussian3 bad mu}\,,\\
    \hat{\sigma}^2 & = (m_{d^2} - m_d^2) - \sigma_e^2\,. \label{eq:gaussian3 bad sigma}
\end{align}
Comparing Eqs.~\eqref{eq:gaussian3 bad mu} and~\eqref{eq:gaussian3 bad sigma} with the expectation value  $E$ and the variance $V$ from Eq.~\eqref{eq:gaussian3 good theta given det}, they find
\begin{align}
    \lim\limits_{N_\mathrm{E}\rightarrow\infty} \left[ \hat{\mu} - \mathrm{E}[\theta]_{p(\theta|{\rm det},\Lambda)} \right] & = \frac{\sigma_e^2}{\sigma_\Lambda^2 + \sigma_e^2 + \sigma_D^2} (\mu_D - \mu_\Lambda)\,, \label{eq:bias in mean} \\
    \lim\limits_{N_\mathrm{E}\rightarrow\infty} \left[ \hat{\sigma}^2 - \mathrm{V}[\theta]_{p(\theta|{\rm det},\Lambda)} \right] & = - \frac{(2\sigma_\Lambda^2 + \sigma_e^2) \sigma_e^2}{\sigma_\Lambda^2 + \sigma_e^2 + \sigma_D^2} < 0\,,
\end{align}
and conclude that such an approach is biased.

%\prlsubsec{Fitting the observed distribution: correct solution}

Let us repeat the last steps, now using the proper Eq.~\eqref{eq:likelihood_pobs} to obtain the posterior. The probability to detect an event $\theta$ is
\begin{align}
    p({\rm det}|\theta)&=\int p({\rm det}|d)p(d|\theta){\rm d} \theta \notag \\
    &= \frac{\sigma_D}{\sqrt{\sigma_D^2+\sigma_e^2}} \exp\left( -\frac{(\theta-\mu_D)^2}{2(\sigma_e^2+\sigma_D^2)}\right)
\end{align}
Renormalizing the likelihood with this single event detection probability, we get 
\begin{align}
    &p(\Lambda|\{d_i,{\rm det}_i\}, N_\mathrm{E}) \nonumber \\ & \propto
    \exp\left\{ -\frac{ \displaystyle \sum_i^N  \left[ \left (\frac{\sigma_e^2+\sigma_D^2}{\sigma_D^2}d_i-\frac{\sigma_e^2}{\sigma_D^2}\mu_D\right )-\mu\right]^2}{\displaystyle 2 \left[ \sigma^2 + \frac{\sigma_e^2}{\sigma_D^2}(\sigma_e^2+\sigma_D^2) \right]} \right\}. 
\end{align}
The estimators for $\mu$ and $\sigma$ become
\begin{align}
    \hat{\mu}&=\frac{\sigma_e^2+\sigma_D^2}{\sigma_D^2}m_d-\frac{\sigma_e^2}{\sigma_D^2}\mu_D,
%\end{align}
%\begin{equation}
\\
\hat{\sigma}&=\left (\frac{\sigma_e^2+\sigma_D^2}{\sigma_D^2}\right)^2(m_{d^2} - m_d^2) - \frac{\sigma_e^2}{\sigma_D^2}(\sigma_e^2+\sigma_D^2)\,,
\end{align}
which, using Eqs.~\eqref{eq:gaussian2 data m1} and \eqref{eq:gaussian2 data m2}, in the limit where $N_\mathrm{E}$ goes to infinity, approach
\begin{align}
\hat{\mu}=&\frac{(\sigma_e^2+\sigma_D^2)\mu_{\Lambda}+\sigma_{\Lambda}^2  \mu_D}{\sigma_{\Lambda}^2+\sigma_D^2+\sigma_e^2}\,, \;\;
\hat{\sigma}=\frac{\sigma_{\Lambda}^2(\sigma_e^2+\sigma_D^2)}{\sigma_{\Lambda}^2+\sigma_e^2+\sigma_D^2}\,.
\end{align}
This is exactly the expectation value $\mathrm{E}[\theta]_{p(\theta|{\rm det},\Lambda)}$ and the variance  $\mathrm{V}[\theta]_{p(\theta|{\rm det},\Lambda)}$ for the detected $\theta$ when fitting the intrinsic distribution, see Eq.~\eqref{eq:gaussian3 good theta given det}. This shows that, with this corrected approach, inference on the observed population is properly unbiased.

\section{Population-synthesis model}\label{app:pop_synth}
The astrophysical population of BBH mergers is constructed leveraging on an upgraded version of the \bpop semi-analytic population synthesis tool \cite{2019MNRAS.482.2991A, 2020ApJ...894..133A, 2023MNRAS.520.5259A,Arca_sedda_prep}. \bpop is a flexible tool that permits the user to simulate a heterogeneous population of BBH mergers originating in different environments, namely in galactic fields---forming via isolated binary evolution---or in young, globular, and nuclear star clusters, where they form via dynamical interactions. We assume that BBHs from isolated binaries and young clusters distribute across cosmic times following the cosmic star-formation history as in Ref. \cite{2017ApJ...840...39M} (see also Refs.~\cite{2022MNRAS.511.5797M,2023MNRAS.520.5259A}). The same star-formation history is adopted for nuclear star clusters, following the idea that these objects form in the centers of their host galaxies and should share, at least partly, their formation history \cite{2020A&ARv..28....4N}. For globular clusters, we instead adopt a Gaussian distribution that evolves with redshift, following \cite{2022MNRAS.511.5797M}. Additionally, the model assumes that the metallicity of the BBH progenitors evolves with redshift following a log-normal distribution peaked around a mean value inferred from cosmological observations \cite{2017ApJ...840...39M,2020A&A...635A..97B,2022MNRAS.511.5797M}. Natal kicks and masses of BBHs are drawn from a stellar evolution catalog assembled with the \textsc{MOBSE} population-synthesis code~\cite{Giacobbo:2017qhh,Giacobbo:2018etu,2020ApJ...891..141G,2023MNRAS.520.5259A}. Spins are assumed to be isotropically distributed in space for dynamically formed BBHs, while for those formed in isolated binaries we follow the prescription for mildly aligned spins of Ref.~\cite{2023MNRAS.520.5259A}. To build the final catalog, we distribute BBH mergers according to the merger rate density calculated for each channel, assuming 1 year of events and a maximum redshift $z=20$~\cite{Arca_sedda_prep}.

%\pagebreak
 \FloatBarrier
\bibliography{Ref}

\end{document}